\documentclass[11pt]{article}

\usepackage{moriond,epsfig}

\def\ra{$\rightarrow$}

\def\be{\begin{equation}}

\def\ee{\end{equation}}

\def\bea{\begin{eqnarray}}

\def\eea{\end{eqnarray}}

\catcode`@=11
\def\chkspace{%
  \relax   
  \begingroup\ifhmode\aftergroup\dochksp@ce\fi\endgroup}
\def\dochksp@ce{%
  \unskip              
  \futurelet\chkspct@k\d@chkspc  
}
\def\d@chkspc{%
  \let\nxtsp@ce=\relax
  \ifx\chkspct@k.\else     
    \ifx\chkspct@k,\else
      \ifx\chkspct@k;\else
        \ifx\chkspct@k!\else
          \ifx\chkspct@k?\else
            \ifx\chkspct@k:\else
              \ifx\chkspct@k)\else
              \ifx\chkspct@k(\else
                \ifx\chkspct@k]\else
                  \ifx\chkspct@k-\else
                    \ifx\chkspct@k\egroup\else  
                      \let\nxtsp@ce=\put@space  
                    \fi
                  \fi
                \fi
              \fi
              \fi
            \fi
          \fi
        \fi
      \fi
    \fi
  \fi
  \nxtsp@ce
}
\def\put@space{$\;$}
\catcode`@=12
\def\ep{{$e^+e^-$}\chkspace}

\def\z0{{$Z^0$}\chkspace}

\def\bb{{$b\bar{b}$}\chkspace}

\def\adhoc{{\it ad hoc}\chkspace}

\begin{document}
\vspace*{4cm}

\title{$B$ Hadron Production and \bb Correlations in \z0 Decays at SLD}

\author{Gavin Nesom}

\address{Nuclear and Astrophysics Laboratory, Oxford University,\\
Keble Road, Oxford, OX1 3RH, England}

\maketitle\abstracts
{
We present results of three SLD analyses: our final determination of the rate 
of gluon splitting into $b\bar{b}$, an improved measurement of the inclusive 
$b$ quark fragmentation function in \z0 decays, and a preliminary first 
measurement of the energy correlation between the two leading B hadrons in 
\z0decays.  
Our results are obtained using hadronic \z0decays produced in \ep 
annihilations 
at the Stanford Linear Collider (SLC) between 1996 and 1998
and collected in the SLC Large Detector (SLD).  In this period, we used 
an upgraded vertex detector with
wide acceptance and excellent impact parameter resolution, 
thus improving considerably our tagging capability for low-energy
$B$ hadrons.
}

\section{Introduction}

The vertex representing a gluon splitting into a \bb pair, $g$\ra\bb
, is a fundamental component of QCD, 
but the contribution of this vertex to physical processes is 
poorly known, both theoretically and experimentally.  
In high-energy \ep annihilation the leading-order process containing
this vertex is $e^+e^-$\ra$q\bar{q}g$\ra
$q\bar{q}b\bar{b}$; 
information on $g$\ra\bb can
thus be obtained by studying \ep\ra hadrons events comprising 
four 
jets,
with two of the jets identified 
as $b$ or $\bar{b}$.  
We define the rate $g_{b\bar{b}}$ as the fraction of $e^+e^-\rightarrow$hadrons
events in which a gluon splits into \bb.  
The measurement of $g_{b\bar{b}}$ is
difficult experimentally
since the rate is intrinsically low and 
the backgrounds from \z0\ra\bb events are two orders of
magnitude larger. In addition, 
the $B$ hadrons from $g$\ra\bb have relatively low energy and  
short flight distance and are difficult to identify using 
standard tagging techniques.  We present a world's best measurement of 
$g_{b\bar{b}}$ using our vertexing ability to find low energy B hadrons.
%

According to the factorization theorem, the heavy quark 
fragmentation function can be described as a convolution of perturbative
and non-perturbative effects.  For the $b$ quark, 
the perturbative calculation is in principle 
understood, 
and the
non-perturbative effects have been parametrized in 
both model-dependent, 
and model-independent approaches.\cite{waffle}  
The most sensitive experimental determination of the fragmentation function 
is expected to come from a precise determination of the $B$ hadron energy 
distribution.
We have previously reported the first measurement of this distribution 
over the entire allowed kinematic range; \cite{ourprl} here we 
update that result and exclude many models and place stringent bounds on the
form of the distribution.

Extending the fragmentation analysis to look at the energy correlation between 
the two B hadrons in a \z0\ra\bb event futher constrains theoretical work.  
It can 
provide a direct test
of the QCD factorisation theorem,\cite{bdh} and further probe phenomological
models.  For example, models that accurately predict the inclusive energy 
distribution might not find the correct correlation, and vice versa.  We present
an analysis that measures the correlation and compares it with a leading order
QCD calculation.


\section{Selecting $B$ Decay Vertices}

A general description of the SLD detector can be found 
elsewhere, 
as can the trigger information and selection criteria
for \z0$\rightarrow$hadron events.\cite{sld1,sld2}  
The $B$ sample is selected using topologically reconstructed secondary vertices
based on the detection and measurement of charged tracks.\cite{zvnim}  
To reconstruct the secondary vertices, the space points 
where track density functions overlap were found in three dimensions.
Only vertices that are significantly displaced from the IP
were considered to be a possible
$B$-hadron.
With SLD's good track impact parameter resolution and precisely know 
interaction point, the
vertex finding gives excellent $b$-tagging efficiency and purity.
In particular, the efficiency is good even at low $B$-hadron energies,
which is especially important for detecting $g$\ra\bb, and for shape 
resolution of the fragmentation function.
The mass of the reconstructed vertex, $M_{ch}$, is calculated by 
assigning each track associated with it, the charged-pion mass.  
Because of the tiny SLC IP error and the excellent 
vertex resolution, 
the transverse momentum $P_t$ of tracks associated with 
the vertex relative to the 
line joining the IP and the secondary vertex is well-measured.
The mass of the missing particles can then 
be partially compensated by using $P_t$ to form the 
``$P_t$ corrected mass'', $M_{P_t} = \sqrt{M_{ch}^2 + P_t^2} + |P_t|$. 

\section{The Rate of Gluon Splitting into $b\bar{b}$ Pairs}




Candidate events containing a gluon splitting into a $b\bar{b}$ pair, 
$Z^0 \!\rightarrow \! q\bar{q}g \!\rightarrow \!q\bar{q}b\bar{b}$,
where the initial $q\bar{q}$ can be any flavor, are required to have 4 jets
(Durham algorithm, $y_{cut}=0.005$).
A secondary vertex is required in each of the two jets 
with the smallest opening angle in the event, yielding 1514 events.
This sample is dominated by background ($S/N \sim 1/10$), primarily from
$Z^0 \rightarrow b\bar{b}g(g)$ events, and events with a gluon splitting
into a $c\bar{c}$ pair.

In order to improve the signal/background ratio, a neural network technique
is used.
Nine observables are chosen for inputs to the neural network.
The inputs include the the $P_T$-corrected masses of the vertices and
the angle between the vertex axes:  
$b$ jets have higher $P_{T}$-corrected mass than $c$/$uds$ jets; 
many $Z^0\to b\bar{b}$ background events have one $b$-jet which
was artificially split by the jet-finder into 2 jets 
that
tend to be collinear.  
The neural network is trained using Monte Carlo samples of 1800k
$Z\to q\bar{q}$ events, 1200k $Z\to b\bar{b}$ events, 780k $Z \to c\bar{c}$
events and 50k $g\to b\bar{b}$ events.
A cut on the neural network output of $> 0.7$ keeps 
79 events in the data, with an estimated background of 41.9 events.  
Using this and the estimated efficiency for selecting
$g\rightarrow b\bar{b}$ splittings of 4.99\% yields a measured fraction of
hadronic events containing such a splitting of: 
\be
g_{b\bar{b}} = (2.44 \pm 0.59 (stat.) \pm 0.34 (syst.)) \times 10^{-3}
\ee
The result is consistent with previous results and the QCD 
prediction~\cite{Miller:1998ig} of $1.75 \times 10^{-3}$.

\section{The Inclusive $b$ Quark Fragmentation Function}

Candidate events for this analysis were found by requiring a secondary 
vertex in an event hemisphere with each hemisphere of the event being 
treated independently.   
Placing the condition that the vertex 
must have an 
$M_{pt} > 2$ GeV and a distance from the I.P. $> 0.1$ yields a B hadron 
sample of 42,093 from the data with an estimated purity of 98.2\%.
Assuming all the tracks associated with the vertex 
are pions we calculate the charged track energy, $E_{ch}$, and the total 
momentum vector $\vec{P}$ of the $B$ vertex; 
we must then find the energy of particles missing from the vertex.  We asssume
$P_{T missing} = P_{T charged}$ and $M_B = 5.26 GeV$ leading to 
an upper bound 
on the missing mass $M_0$:  
$M_{0max}^2 = M_B^2 - 2M_B\sqrt{M_{ch}^2 + P_t^2} + M_{ch}^2$.  
The true missing mass $M_0^{true}$ is 
often rather close to $M_{0max}$, 
and we use the latter 
as an estimate of $M_0^{true}$.
We can then solve
 for the longitudinal momentum of the missing particles using kinematic 
considerations, 
and hence the missing energy from the vertex, $E_0$.  
The $B$ hadron energy is then $E_B = E_{ch} + E_0$.
Since $ 0 \leq  M_0^{true} \leq M_{0max}$, 
the $B$ energy is well-constrained when $M_{0max}$ is small.  
For this reason, in order to obtain good energy resolution, we cut events from 
our sample using a value of $M_{0max}^2$ that depends on $E_B$ to achieve 
a $B$ selection efficiency nearly independant of $x_B$.
A total of 4,164 vertices were found after all cut criteria, with an 
estimated purity of 99.0\%.  The full kinematic range is covered by 
these events.



\begin{figure}[h]
\center\psfig{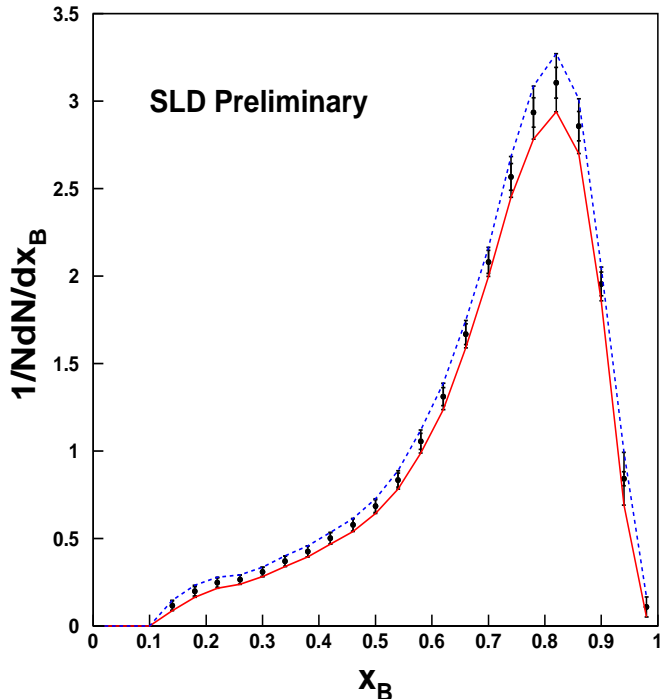}
\vskip 0.2cm
\caption{The corrected single inclusive weakly deacying B hadron energy 
distribtion.  The points represent the data, and the envelope is our total 
error.
\label{fig:unfold}}
\vskip -0.5cm
\end{figure}

After background subtraction, 
the distribution of the reconstructed scaled $B$ hadron energy  
is compared with several heavy quark fragmentation models.
Within the context of the JETSET Monte Carlo the 
Bowler,
Lund,
and Kartvelishvili 
models are consistent with the data as is the stand-alone UCLA generator.
We also compared the data with a set of {\em ad hoc} functional forms of the 
$x_B$ distribution in order to estimate the variation 
in the shape of the $x_B$ distribution. 
Of those we tested, only the Peterson function
, two \adhoc 
generalizations of the 
Peterson function 
and an 8th-order 
polynomial
are consistent with the data.\cite{ourprl}  
We use the eight consistent fragmentation models and functional forms 
to unfold the raw data distribution for selection 
efficiency and bin migrations.  Their average is taken as the central 
value in each bin and the 
RMS value as the model dependent error.  The fully corrected data are shown in 
Figure~\ref{fig:unfold}.

\section{The Two Dimensional $B\bar{B}$ Energy Distribution and Correlation}

\begin{figure}[h]
\center\psfig{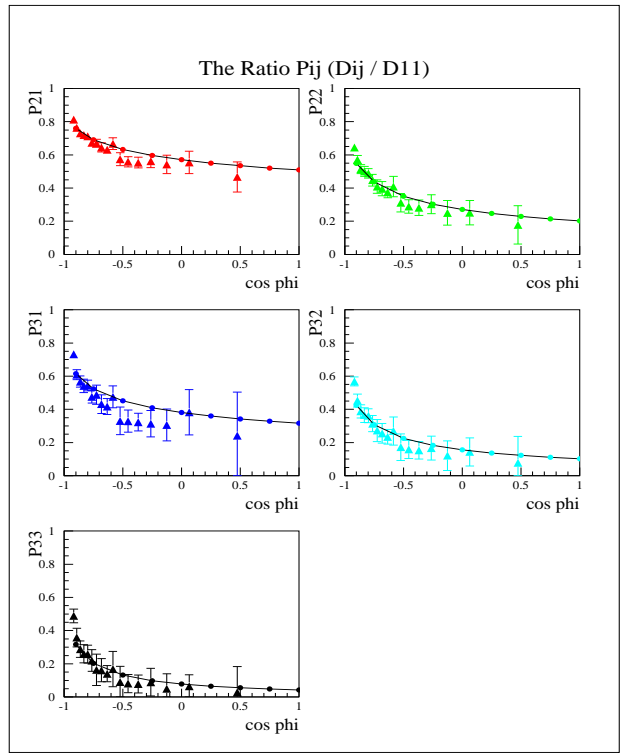}
\vskip -0.2cm
\caption{Ratio of the double moments $P_{ij} / P_{11}$ of the 2-dimensional
B hadron energy distribution.  Data are shown as points with stat. error bars 
only.  A leading order QCD calculation of the ratios are shown as solid lines.
\label{fig:corr}}
\vskip -1.0cm
\end{figure}

The previous analysis has placed considerable constraints on the form
of the $b$ fragmentation function, and on phenomological models.  SLD, 
however, can go further and look at the correlation between the 2 leading 
B hadrons.  Events are selected by requiring that exactly two secondary 
vertices are found in different jets (Durham algorithm, $y_{cut}=0.015$).  
Either vertex is required to have 
an $M_{pt} > 2$ GeV and the same (or otherwise) vertex, is required to be 
seperated from the I.P. by $> 0.1$ cm.  An upper limit on $M_0$ of the 
remaining sample ensures an energy resolution better than 20\%.  A cut
on the angular separation between the two decay vertices avoids 
the problem that sometimes a genuine vertex is split into two separate 
candidates.  After all selection criteria we have 19283 events in the data 
in which 
both the $B$ and $\bar{B}$ are reconstructed with an estimated purity and 
efficiency of 99.6\% and 36.6\% respectively.  We group these events in bins of 
$cos\phi$ where $\phi$ is the angle between the two vertex axes.  
By taking moments of the 2-dimensional energy distribution we have a measure 
of the correlation between the two hadrons.
Within each bin we calculate the double moments $D_{ij} (i,j = 1,2,3)$.  
We use our Monte Carlo to correct the data for selection efficiencies and 
detector effects.  In order to compare with the QCD calculation we take a 
ratio and calculate normalise the $D_{ij}$ to $D_{11}$.  The ratios of these 
moments, $P_{ij}$ are compared with a leading order QCD calculation in 
Figure~\ref{fig:corr}.

\section{Conclusion}

We have presented the results of three analyses, all of which make use of SLD's
exceptional B-tagging performance.
Our final determination of the rate of gluon splitting into $b\bar{b}$ pairs 
per hadronic \z0~decay, $g_{b\bar{b}}$, is a world's best:
$g_{b\bar{b}} = (2.44 \pm 0.59 (stat.) \pm 0.34 (syst.)) \times 10^{-3}$.  
We have updated our measurement of the inclusive weakly decaying B hadron 
energy distribution and exclude many models of b quark fragmentation.  We 
obtain 
a precise measurement of the average scaled B hadron energy:
$<x_B> = 0.710\pm 0.003 (stat)\pm0.005 (syst)\pm0.004 (model)$.  
We have a first measurement of the energy correlation between the leading B 
hadrons in $Z^0$ decays, the results of which are consistent with a leading 
order QCD 
calculation.


\end{document}